\documentclass{article}

\usepackage{amsmath, amssymb}
\usepackage{natbib}
\usepackage{booktabs}
\usepackage{multirow}
\usepackage{tabularx}
\usepackage{hyperref}
\usepackage{authblk}

\hypersetup{pdfstartview = {XYZ null null 1.0}}

\newtheorem{definition}{Definition}

\newcolumntype{C}{>{\centering\arraybackslash}X}
\newcolumntype{L}{>{\raggedright\arraybackslash}X}
\newcolumntype{R}{>{\raggedleft\arraybackslash}X}

\begin{document}

\title{Bayesian Methods to Improve The Accuracy of Differentially Private Measurements of Constrained Parameters}

\author[1]{Ryan Janicki}
\author[2, 3]{Scott H.  Holan}
\author[1]{Kyle M. Irimata}
\author[1]{James Livsey}
\author[1]{Andrew Raim}

\affil[1]{Center for Statistical Research and Methodology, U.S. Census Bureau,
            4600 Silver Hill Road, 
            Washington,
            DC, 20233, 
            USA}
\affil[2]{Department of Statistics, University of Missouri,
            146 Middlebush Hall, 
            Columbia,
            MO,
            65211, 
            USA}
\affil[3]{Office of the Associate Director for Research and Methodology, U.S. Census Bureau,
            4600 Silver Hill Road, 
            Washington,
            DC,
            20233, 
            USA}
            
\maketitle

\begin{abstract}
Formal disclosure avoidance techniques are necessary to ensure that published data can not be used to identify information about individuals. The addition of statistical noise to unpublished data can be implemented to achieve {\em differential privacy}, which provides a formal mathematical privacy guarantee.  However, the infusion of noise results in data releases which are less precise than if no noise had been added, and can lead to some of the individual data points being nonsensical.  Examples of this are estimates of population counts which are negative, or estimates of the ratio of counts which violate known constraints.  A straightforward way to guarantee that published estimates satisfy these known constraints is to specify a statistical model and incorporate a prior on census counts and ratios which properly constrains the parameter space.  We utilize rejection sampling methods for drawing samples from the posterior distribution and we show that this implementation produces estimates of population counts and ratios which maintain formal privacy, are more precise than the original unconstrained noisy measurements, and are guaranteed to satisfy prior constraints.
\end{abstract}

\section{Introduction}\label{sec:intro}

National statistical agencies such as the U.S. Census Bureau are tasked with
collecting information about the population, housing characteristics, and business
establishments, and disseminating data products for public use.  An important
consideration that needs to be made when deciding which datasets can be released is
protection of individual respondent data.  Federal law requires all collected data
to be kept confidential.  In particular, Title 13 of the U.S. Code prohibits the
U.S. Census Bureau from disclosing any ``information reported by, or on behalf of,
any particular respondent'' and from creating ``any publication whereby the data
furnished by any particular establishment or individual under this title can be
identified.'' \footnote{Title 13
U.S. Code, Sections 8--9; Title 13 U.S. Code, Section 141.}
\citep{us2021disclosure}.

Disclosure avoidance (DA) is the process of protecting the confidentiality of
an individual respondent's personal data.  Historically, statistical agencies have used
ad hoc methods such as partial or full data suppression of some data tables
for small geographic areas to avoid indirect disclosure.  More sophisticated methods,
such as data swapping, which swaps records for some households with similar
characteristics, along with top and bottom coding of sensitive data, have also been
used.  See \citet{mckenna2018disclosure} for a historical review of DA techniques
used by the U.S. Census Bureau.

With easy access to increasingly powerful computing resources and advanced data
science tools it has become easier to reconstruct and re-identify information about
individuals from aggregated public-use data products.  Outside parties can now
conceivably take published tables from the U.S. Census Bureau and link to
information from commercial databases.  This database reconstruction has
necessitated more sophisticated DA approaches.  See \citet{mckenna2019us} for
examples of successful re-identification attacks using publicly available data.

Recently, in a series of papers beginning with \citet{dwork2006differential}, a new
DA framework based on {\em differential privacy} (DP) has been developed.  Much of
the appeal of DP is that it gives a rigorous definition of what is meant by privacy
and provides mathematically provable guarantees of privacy protection.  Roughly
speaking, a privacy protection mechanism that provides differential privacy ensures
that the addition or removal of a single record from a database does not
dramatically affect the outcome of any analysis \citep{dwork2008differential}.  DP
has quickly become the gold standard for privacy protection algorithms and has been
adopted by the U.S. Census Bureau for its 2020 decennial data products
\citep{us2021disclosure}.  A comprehensive review of DP can be found in
\citet{dwork2014algorithmic} and a review of historical privacy protection methods at
the U.S. Census Bureau as well as an overview of methods used in the 2020 decennial
census can be found in \citet{abowd2023confidentiality}.

While DP represents a great advancement in the field of disclosure avoidance, it also
introduces serious consequences from the perspective of data users.  With any DA
methodology there is an inherent trade-off between privacy protection and the utility
of data.  DP, in particular, makes use of noise infusion into confidential data to ensure privacy, which
results in published data products that are necessarily less precise than if no DA
mechanism had been employed.  Noise infusion also has the unfortunate effect of causing
estimates to sometimes violate known constraints.  For example, DP can result in
estimates of population counts which are negative, and estimates of ratios which violate logical
constraints.  Recent work on addressing issues with bounded data protected by differential privacy can be found in \citet{kazan2024bayesian}.

In this article we propose a Bayesian model-based solution for producing estimates which are
guaranteed to satisfy all known constraints, which are more precise than the original
noisy measurements, and which maintain all privacy protections guaranteed by the
original DA algorithm.  We model the noisy measurements after implementation of
DP, using the known noise infusion mechanism, and incorporate all knowledge about
logical constraints into the prior distribution on population counts.  We provide a brief overview of
differential privacy in Section~\ref{sec:background} and describe the main results in
Section~\ref{sec:main}.  A data example using 2010 decennial census tables is given
in Section~\ref{sec: example}.  Concluding remarks are made in Section~\ref{sec:conclusion}.

\section{Background}\label{sec:background}

Let $\boldsymbol{X} \subset \mathcal{X}$ denote a dataset (or database) which is a
collection of individual records obtained through a survey sample or a census, and
let $x_i$ be the number of records in $\boldsymbol{X}$ of type $i$.  For
example, $\boldsymbol{X}$ might contain information on a collection of records'
age, race, sex and geography.  Then $i$ would index the possible age by race by sex cross
classifications at each geographic region, and $x_i$ would be the number of individuals in $\boldsymbol{X}$
which belong to cell $i$.  

The statistical agency that collected the data $\boldsymbol{X}$ would like to
release tabulations (or queries) of $\boldsymbol{X}$ to the public, such as the
number of persons at different geographies with certain characteristics of interest,
but to also ensure that information about any one record in $\boldsymbol{X}$ can
not be identified.  The privacy of individual records can be achieved by applying DA
techniques to either $\boldsymbol{X}$ or tabulations of $\boldsymbol{X}$ prior to
dissemination.  DA which satisfies differential privacy is a
principled way to ensure protection of individual respondents.

Differential privacy, which was introduced in \citet{dwork2006differential}, requires
a definition of distance between databases.  For any two databases $\boldsymbol{X}$
and $\boldsymbol{X}'$ in $\mathcal{X}$ we can define their distance using the
$l^1$ norm
\begin{equation}
  \left\| \boldsymbol{X} - \boldsymbol{X}' \right\|_1 = \sum^{\mid \mathcal{X}
    \mid}_{i = 1} \left| x_i - x'_i \right|,
\end{equation}
where $\left| \mathcal{X} \right|$ denotes the number of unique records in
$\mathcal{X}$.  Roughly speaking, if an algorithm provides privacy protections,
then the outputs should be similar when applied to similar databases, so that any
one individual record is not overly influential and can not easily be recovered.  The
following formal definition of differential privacy can be found in
\citet{dwork2014algorithmic}.
\begin{definition}
  A randomized algorithm $\mathcal{M}$ is $(\varepsilon, \delta)$-differentially
  private if for all $\mathcal{S} \subseteq \text{Range} \left( \mathcal{M}
  \right)$ and for all databases $\boldsymbol{X}, \boldsymbol{X}'$ such that
  $\left\| \boldsymbol{X} - \boldsymbol{X}' \right\| \leq 1$,
  \begin{equation*}
    P \left( \mathcal{M} (\boldsymbol{X}) \in \mathcal{S} \right) \leq \exp\{
      \varepsilon \} P \left( \mathcal{M} (\boldsymbol{X}') \in \mathcal{S} \right) +
      \delta,
  \end{equation*}
  for $\varepsilon > 0$ and $\delta \geq 0$.
\end{definition}
The privacy budget is given by the parameters $\varepsilon$ and $\delta$ and
determines the amount of privacy guarantee.  Small values of $\varepsilon$ and
$\delta$ provide greater privacy protection at the expense of less accurate
published data, while larger values of $\varepsilon$ and $\delta$ result in more
accurate data in exchange for weaker privacy guarantees.

Let $Y$ be a tabulation of $\boldsymbol{X}$ that a statistical agency would like
to publish.  For example, $Y$ could represent the number of households by
relationship for the population under 18 years in a county in the U.S.  Additional
examples of tabulations published by the U.S. Census Bureau are given in Section~\ref{sec: example}.  The tabulation $Y$ cannot be released to the public without
first applying DA techniques.  A simple privacy protection algorithm which achieves
differential privacy is adding statistical noise to $Y$ and releasing this noisy
version of $Y$.  The noisy measurement is denoted by $Z$, and can be generated as
\begin{equation}\label{E: NM}
  Z = Y + \varepsilon,
\end{equation}
where $\varepsilon$ is sampled from a noise-generating (probability) distribution.  Two of the
most used distributions for $\varepsilon$ are the Gaussian distribution, which has
probability density function
\begin{equation*}\label{E: Gauss}
  f \left( x; \sigma^2 \right) = \frac{1}{\sqrt{2 \pi \sigma^2}} e^{-\frac{1}{2
    \sigma^2} x^2}
\end{equation*}
and the Laplace distribution, which has density function
\begin{equation*}\label{E: Laplace}
  f \left( x; \lambda \right) = \frac{1}{2 \lambda} e^{-\frac{\mid x
      \mid}{\lambda}}. 
\end{equation*}
The Laplace mechanism applied in this way preserves $(\varepsilon, 0)$-differential
privacy while the Gaussian mechanism preserves $(\varepsilon, \delta)$-differential
privacy \citep{dwork2014algorithmic}.

The addition of statistical noise from a Laplace or Gaussian distribution guarantees
that differential privacy will be satisfied.  However, it has the unfortunate effect
of reducing the utility of the data as the noisy measurements, $Z$, are less
precise than the tabulations $Y$.  The noisy measurements may also violate certain
constraints that the unperturbed tabulations are known to satisfy.  For example, if
$Y$ is a count tabulation, then $Y$ must be nonnegative.  If $Y$ is the ratio
of two count tabulations, there may be a relationship between the numerator and the
denominator that must be taken into account.  Publishing these noisy measurements as
official data products could affect how users interpret the data and may result in
lowered confidence in the quality of the estimates.  In the next section we introduce
model-based methods for improving the quality of noisy (differentially private) measurements obtained by
adding statistical noise to tabulations by incorporating constraints into a prior
distribution.

\section{Modeling set up}\label{sec:main}

Let $\boldsymbol{Y} \in \mathbb{R}^m$ be a vector of tabulations of the database
$\boldsymbol{X}$, and let $\boldsymbol{Z}$ be a privacy-protected measurement of
$\boldsymbol{Y}$ obtained by independently adding noise to each component of
$\boldsymbol{Y}$.  Let $f$ denote the noise generating distribution.  Then
\begin{equation}\label{E:likelihood}
  \boldsymbol{Z} \mid \boldsymbol{Y}, \boldsymbol{\theta} \sim \prod^m_{i = 1} f
    \left( Z_i ; Y_i, \boldsymbol{\theta} \right)
\end{equation}
where $\boldsymbol{\theta}$ is a vector of parameters determined by the DA
algorithm which will be fully known to the analyst.  In many situations we will have
prior knowledge about logical constraints that must be satisfied by the vector
$\boldsymbol{Y}$, but that are not necessarily respected by the vector of noisy
measurements, $\boldsymbol{Z}$.  For example, if $\boldsymbol{Y}$ is a vector of
counts, it follows that each component must be nonnegative.  Another potential issue
occurs in tables consisting of ratios.  Some examples are given in Section~\ref{sec:
  example}.

Let $p$ denote the number of known inequality constraints that must be satisfied by
the components of $\boldsymbol{Y}$.  We can summarize this information in terms of
a vector of lower bounds, $\boldsymbol{l} \in \mathbb{R}^p$, a vector of upper
bounds, $\boldsymbol{u} \in \mathbb{R}^p$, and a constraint matrix $\boldsymbol{D}
\in \mathbb{R}^{p \times m}$,
\begin{equation}\label{E:constraints}
  \boldsymbol{l} \leq \boldsymbol{D} \boldsymbol{Y} \leq \boldsymbol{u},
\end{equation}
where the inequalities are to be interpreted componentwise.  A straightforward way
to incorporate the constraints is to use a prior distribution on $\boldsymbol{Y}$ with support implied
by the inequalities in \eqref{E:constraints}.  For our work we used an improper
distribution
\begin{equation}\label{E:prior}
  \pi \left( \boldsymbol{Y} \right) \propto I \left( \boldsymbol{l} \leq
    \boldsymbol{D} \boldsymbol{Y} \leq \boldsymbol{u} \right),
\end{equation}
where $I (\cdot)$ is the indicator function.  Combining \eqref{E:likelihood} and
\eqref{E:prior} results in a posterior distribution
\begin{equation}\label{E:posterior}
  \boldsymbol{Y} \mid \boldsymbol{Z}, \boldsymbol{\theta} \propto \prod^m_{i = 1} f
    \left( Z_i ; Y_i, \boldsymbol{\theta} \right) I \left( \boldsymbol{l} \leq
    \boldsymbol{D} \boldsymbol{Y} \leq \boldsymbol{u} \right).
\end{equation}
Since the prior is improper when either the upper or lower bound is infinite, it does need to be verified that the expression in
\eqref{E:posterior} is integrable.  Fortunately, in most practical applications, the
noise will be additive so that $f$ is location invariant and \eqref{E:posterior}
will be proper.

Samples drawn from \eqref{E:posterior} will be guaranteed to satisfy
\eqref{E:constraints}.  The problem becomes more of a computational one as it is not
straightforward to sample from \eqref{E:posterior} efficiently.  Neither the
posterior distribution \eqref{E:posterior} nor any of its full conditional
distributions belong to standard parametric families for any choice of $f$, so
there is no known way to directly sample from \eqref{E:posterior}.  Instead, indirect
methods involving proposal distributions and accept/reject algorithms such as the
Metropolis Hastings method \citep{metropolis1953equation} must be used.  Gibbs
sampling under inequality constraints was studied in \citet{gelfand1992bayesian}, and
can be done with univariate cross-sectional Gibbs sampling which draws from the
unconstrained model and only retains samples which lie in the constrained space.
Another approach is to sample from the unrestricted posterior distribution and either
reject samples which fall outside the constrained space, or project onto the
constrained space \citep{dunson2003bayesian}.  The main issue with these methods is
efficiency as accept/reject algorithms can result in most proposals being rejected if
not properly tuned.

For the special case when $f$ is Gaussian, the posterior distribution is a
multivariate truncated Gaussian distribution.  Sampling from a multivariate truncated
Gaussian distribution has been studied by several authors, including
\citet{li2015efficient}, \citet{ma2020sampling}, \citet{ghosal2022bayesian}.  An
efficient implementation of the algorithm in \citet{ma2020sampling} is in the R
package \texttt{tmvmixnorm} \citep{ma2020sampling}.

We are not aware of custom samplers for other multivariate truncated distributions.
If the noise mechanism is some other distribution, we propose using a multivariate
truncated Gaussian that is close to the target distribution as a proposal.  The
Gaussian distribution can be chosen by a variety of methods such as moment matching
or minimizing a distance.  When a Laplace distribution is used, there is a simple,
closed-form expression which minimizes the Kullback-Leibler divergence from a
Gaussian distribution.  Using an approximating Gaussian distribution was found to be
an effective solution in \citet{irimata2022evaluation}.

Samples generated from \eqref{E:posterior} are forced to satisfy all logical
constraints and estimates.  Importantly, model estimates of $\mathbf{Y}$ made using the
posterior samples (e.g. the posterior mean) will also maintain all privacy protection
guarantees as the noisy measurements, $\boldsymbol{Z}$.  Using posterior inference
for estimation of $\boldsymbol{Y}$ can be thought of as a postprocessing of the
noisy measurements, and postprocessing maintains the same level of privacy protections as the preprocessed noisy measurements \citep[Proposition 2.1]{dwork2014algorithmic}.

In addition, the posterior mean will generally be more precise than the noisy
measurements as we are utilizing additional information about the unknown parameter
$\mathbf{Y}$ through the constraints in the prior distribution.  Note that because
the noise generating mechanism, including all parameters, is completely known, and
the constraints are based on accurate prior information, there is no possibility of
model misspecification.  The constraints imposed in the prior effectively reduce the
parameter space.  Intuitively, if it is known that the true data-generating process
is a submodel of some working model, then precision should be increased by using the
submodel rather than the full model.  An asymptotic result supporting this idea can
be found in \citet{altham1984improving}.  We also verify this empirically in the next
section.

\section{Example}\label{sec: example}

This project was motivated by the production needs of the Social, Economic, and
Housing Statistics (SEHSD) and Population (POP) Divisions of the U.S. Census Bureau.
SEHSD and POP produce the Supplemental Detailed Housing Characteristic (S-DHC) tables
which contain information about characteristics of persons, households, and
person-household joins (tables which combine person data and household data).  There are 8 tables included in the S-DHC.  These tables are
Average Household Size by Age (PH1), Household Type for the Population in Households
(PH2), Households by Relationship For the Population Under 18 years (PH3), Population
in Families by Age (PH4), Average Family Size by Age (PH5), Family Type and Age For
Own Children Under 18 years (PH6), Total Population in Occupied Housing Units by
Tenure (PH7), and Average Household Size of Occupied Housing Units by Tenure (PH8).
These tables will be published at the nation and state levels of geography using 2020
census data, and 6 of the eight tables will be iterated by major race category and
Hispanic origin.  Further detail about S-DHC and other 2020 decennial census data
products can be found at\\
\url{https://www.census.gov/programs-surveys/decennial-census/decade/2020/planning-management/process/disclosure-avoidance/newsletters/update-2020-census-data-products.html}.

\begin{table}[t]
\centering
\begin{tabularx}{\textwidth}{l l C C C C C C}
  \toprule
  &  & Alabama & Alaska & Arizona & Arkansas & California \\ 
  \midrule
    Total: &                    & 3.02 & 3.21 & 3.19 & 3.00 & 3.45 \\
           & Under 18 years:    & 0.87 & 1.07 & 1.01 & 0.90 & 1.05 \\
           & 18 years and over: & 2.14 & 2.14 & 2.18 & 2.10 & 2.40 \\
  \bottomrule
\end{tabularx}
\caption{Average family size by age:  2010 published decennial census state-level total population tabulations.  Available at \url{data.census.gov}. \label{tab: p37}}
\end{table}

In this section we give an example using the 2010 version of the PH5 table, Average
Family Size by Age, Race and Ethnicity in states in the U.S.  The race and
ethnicity iterations are White alone; Black or African American alone; Asian alone; American Indian and Alaska Native alone; Native Hawaiian and Other Pacific Islander alone; Some Other Race alone; Two or More Races; Hispanic or Latino; White alone, not Hispanic or Latino; and unattributed.  The estimates that are produced in this table
are the ratio of number of persons 18 and under in families to the number of family
households, the number of persons over 18 in families to the number of family
households, and the total number of persons in families to the number of family
households.  Table \ref{tab: p37} shows the published 2010 state-level ratios for total population for five states.

Three noisy measurements are generated for each geography and each race iteration for
the PH5 table: the total population under 18 in families, the total population over
18 in families, and the number of family households.  Denote the true counts as
$Y_{18-}$, $Y_{18+}$, and $Y_\text{FHH}$, and the noisy measurements as $Z_{18-}$,
$Z_{18+}$, and $Z_\text{FHH}$.  The published values in PH5 are
\begin{equation}\label{E:ratios}
  \frac{Z_{18-}}{Z_\text{FHH}}, \ \frac{Z_{18+}}{Z_\text{FHH}}, \text{ and } \frac{Z_{18-} +
    Z_{18+}}{Z_\text{FHH}},
\end{equation}
which are estimates of the ratios \
\begin{equation}\label{E:truth}
  \frac{Y_{18-}}{Y_\text{FHH}}, \ \frac{Y_{18+}}{Y_\text{FHH}}, \text{ and } \frac{Y_{18-} +
    Y_{18+}}{Y_\text{FHH}},
\end{equation}
respectively.

The constraints that must be satisfied are $Y_{18-} \geq 0$, $Y_{18+} \geq
0$, $Y_\text{FHH} \geq 1$ (we only consider areas with at least one occupied
household), and $Y_{18+} + Y_{18-} \geq 2 Y_\text{FHH}$ (the universe is family
households).  We have an additional constraint that is specific to our application
that $Y_{18+} + Y_{18-} \leq \kappa Y_\text{FHH}$, where $\kappa$ is a positive
integer.  This constraint is due to the privacy algorithm used by the U.S. Census
Bureau for person-household join tables which truncates the family household universe
to households with at most $\kappa$ individuals.  Let $\boldsymbol{Y}^\intercal =
\left( Y_{18-}, Y_{18+}, Y_\text{FHH} \right)$ and $\boldsymbol{Z}^\intercal = \left(
Z_{18-}, Z_{18+}, Z_\text{FHH} \right)$.  For this problem, the constraints in
\eqref{E:constraints} are
\begin{equation}\label{E:ph5const}
  \boldsymbol{l} = \begin{bmatrix} 0 \\ 0 \\ 1 \\ 0 \\ 0 \end{bmatrix}, \ 
  \boldsymbol{D} = \begin{bmatrix} 1 & 0 & 0 \\
                                   0 & 1 & 0 \\
                                   0 & 0 & 1 \\
                                   1 & 1 & -2 \\
                                   -1 & -1 & \kappa \\
                   \end{bmatrix}, \
  \boldsymbol{u} = \begin{bmatrix} \infty \\ \infty \\ \infty \\ \infty
    \\ \infty \end{bmatrix}.
\end{equation}

When all of the noisy measurements $Z_{18-}$, $Z_{18+}$ and $Z_\text{FHH}$ are large
relative to the amount of noise added, the ratios in \eqref{E:ratios} will typically
be sensible, accurate estimates of the true ratios in \eqref{E:truth}.  However, when
one or more of the true counts which make up the true ratios in \eqref{E:truth} is
very small, the noisy measurements of these counts can be negative, resulting in a
ratio which is negative.  Furthermore, if $Y_\text{FHH}$ is close to zero, the ratios of
noisy measurements ``blow up'' and appear as an impossibly large number.

The preliminary DA algorithm uses a privacy-loss budget which results in 90\% margins
of error of 200 and a truncation level of 10.  This 90\% margin of error is
equivalent to a variance parameter of $\sigma^2 = \text{14,782}$ when using a Gaussian noise distribution,
or a scale parameter of $\lambda = 86.86$ when using a Laplace noise distribution.  We performed
two experiments based on these parameter settings.  We first generated a set of noisy
measurements by adding independent Gaussian noise with variance $\sigma^2 = \text{14,782}$ to the true
2010 census counts described above.  We then drew 10,000 samples from the posterior
distribution \eqref{E:posterior} using the correctly-specified Gaussian likelihood
and constraints as in \eqref{E:ph5const}.  All computational work was done using
\texttt{R} \citep{r2021r} and samples were drawn from the multivariate truncated
normal distribution using the \texttt{rtmvn} function in the \texttt{tmvmixnorm}
package \citep{ma2020sampling}.

We then repeated this experiment, but instead added independent Laplace noise with
the scale parameter set to 86.86.  The posterior distribution in \eqref{E:posterior}
is then a truncated multivariate Laplace distribution.  Since there is no way to
directly sample from this distribution, we instead used the Metropolis-Hastings
algorithm to draw samples \citep{metropolis1953equation}.  We used a multivariate
truncated Gaussian distribution as the proposal distribution with the variance set to
$\sigma^2 = \pi 86.86^2 / 2$; it is easy to verify that this minimizes the
Kullback-Leibler distance to the truncated multivariate Laplace distribution
\citep{irimata2022evaluation}.

\begin{table}[ht]
\centering
\begin{tabularx}{\textwidth}{l C C C C C C C}
  \toprule
  Mechanism & Estimate & MIN & MAX & BAD\% & RMSE & COV & LEN \\
  \midrule
  \multirow{2}{*}{Gauss} & NM & -5.2 & 17.2 & 1.4 & 0.7 & 87.5 & 1.2 \\
                         & MB & 0.5  &  6.2 & 0.0 & 0.2 & 89.3 & 0.3 \\
  \midrule
  \multirow{2}{*}{Laplace} & NM & -6.6 & 12.5 & 1.4 & 0.6 &   NA & NA \\
                           & MB & 0.5  &  6.2 & 0.0 & 0.2 & 86.7 & 0.3 \\
  \bottomrule
\end{tabularx}
\caption{Comparison of Noisy Measurements (NM) and Model-based predictions (MB) when the
  noisy measurements are generated using either a Gaussian mechanism or a Laplace
  mechanism.  The metrics shown are the maximum value (MAX), minimum value (MIN),
  the percent which are outside the constrained region (BAD\%), root mean squared
  error (RMSE), coverage rate (COV) and  interval length (LEN). \label{tab: results}}
\end{table}

Table \ref{tab: results} presents summary measures of the noisy measurements and the
model-based estimates for each experiment.  Recall that the true value of each ratio
must be between 0 and 10.  The metrics from the 10,000 samples are: the minimum value of
all ratios (MIN), the maximum value of all ratios (MAX), the percent of the estimates
which violate the constraints in \eqref{E:ph5const} (BAD\%), the root mean squared
error of the ratios (RMSE), the average coverage rate of the interval estimates
(COV) and the average length of the interval estimates (LEN).

When the noisy measurements are generated from the Gaussian distribution, confidence
intervals for the ratio of noisy measurements can be calculated using Fieller's method
\citep{fieller1954some}.  There is no analogous method for obtaining confidence
intervals for the ratio of Laplace variables, so the COV and LEN columns for the
noisy measurements for this experiment are omitted.  The interval estimates for the
predicted values were taken to be the 5th and 95th percentiles of the samples taken
from the posterior distribution.

In both experiments, only 1.4\% of the ratios violated the constraints in
\eqref{E:ph5const}.  However, the consequences of these violations become clear when
looking at the results in Table \ref{tab: results}.  In both the Gaussian and Laplace
experiments there are estimates of average number of persons in family households at
the state level that are either negative or unreasonably large.  Publishing such
numbers could result in lack of confidence from data users.  The model-based
post-processing of the noisy measurements eliminated all nonsensical ratio estimates
and drastically reduced the root mean squared error and length of the coverage
intervals.  Also, the MH algorithm used with the Laplace mechanism had a high
acceptance rate and the precision of the ratios was as good as the experiment using
the Gaussian mechanism, although the coverage rate of the intervals was slightly
below the nominal rate.

\section{Conclusion}\label{sec:conclusion}

The analysis shown in Section~\ref{sec: example} was for a person-household join
table which requires a rather large privacy budget compared to other decennial census
products, and also requires a truncation of the housing universe to households
containing a pre-specified maximum number of persons.  We gave an example which
produced a set of noisy measurements for this table at the U.S. state level, and
post-processed these noisy measurements using a model through posterior inference using a prior
distribution which incorporates known constraints.  We demonstrated that this model-based
procedure results in estimates which are more precise than the noisy measurements and
belong to the constrained parameter space.

In the future we would like to produce estimates at substate geographies, such as
county, tract, and block group levels.  Generating noisy measurements for these
geographies requires a much larger privacy budget and would result in ratios which
more often violate the constraints.  Future research is needed to determine whether this procedure
still results in publishable estimates at this higher level of noise.  Future work is
also needed to determine whether covariate information can be utilized to further improve the
quality of estimates.

\section*{Acknowledgments}
  This article is released to inform interested parties of ongoing research and to encourage discussion. The views expressed on statistical issues are those of the authors and not those of the U.S. Census Bureau.

\bibliographystyle{cas-model2-names}

\bibliography{refs}

\end{document}